\documentclass[preprint,12pt,3p,twocolumn]{elsarticle}

\usepackage{graphicx}
\usepackage{amssymb}
\usepackage{lineno}
\usepackage[utf8]{inputenc}
\usepackage{cleveref}
\crefname{section}{§}{§§}
\Crefname{section}{§}{§§}

\journal{journal}

\begin{document}
\begin{frontmatter}

\title{Measuring the temperature and diversity of the \\ U.S. regulatory ecosystem}



\author[iit,codex]{Michael J Bommarito II}
\author[iit,codex]{Daniel Martin Katz}
 
\address[iit]{Illinois Tech - Chicago Kent College of Law}
\address[codex]{CodeX - The Stanford Center for Legal Informatics}

\begin{abstract}
Over the last 23 years, the U.S. Securities and Exchange Commission has required over 34,000 companies to file over 165,000 annual reports.  These reports, the so-called ``Form 10-Ks,'' contain a characterization of a company's financial performance and its risks, including the regulatory environment in which a company operates.  In this paper, we analyze over 4.5 million references to U.S. Federal Acts and Agencies contained within these reports to build a mean-field measurement of temperature and diversity in this regulatory ecosystem, where companies are organisms inhabiting the regulatory environment.  While individuals across the political, economic, and academic world frequently refer to trends in this regulatory ecosystem, far less attention has been paid to supporting such claims with large-scale, longitudinal data. In this paper, we document an increase in the regulatory energy per filing, i.e., a warming ``temperature.''  We also find that the diversity of the regulatory ecosystem has been increasing over the past two decades, as measured by the dimensionality of the regulatory space and distance between the ``regulatory bitstrings'' of companies.  These findings support the claim that regulatory activity and complexity are increasing, and this measurement framework contributes an important step towards improving academic and policy discussions around legal complexity and regulation.
\\
\end{abstract}

\begin{keyword}
complex systems \sep natural language \sep temperature \sep diversity \sep legal complexity \sep regulation
\end{keyword}

\end{frontmatter}


\section{Introduction}
\label{S:1}
Economies, like ecosystems, exhibit dynamic, complex behaviors resulting from the interaction of ``organisms'' inhabiting and altering their ``environment.''  In the case of economies, organisms can be seen as companies, and environments can be seen, at least in part, as regulations.  Just as changes in the environment like rising temperature can harm or help organisms, either broadly or for specific regions or organisms, so too can regulation harm or help companies.  Yet unlike studies of biological ecosystems, studies of the economy have thus far lacked a longitudinal, empirical measure of fundamental environment factors like ``temperature'' or ``species diversity.''  In this paper, we attempt to bridge this gap, finding support for the common claim that regulatory activity and complexity has increased over the last 20 years.

Each year, companies that meet the registration requirements under the Securities and Securities Exchange Acts of 1933 and 1934 must file an annual report with the U.S. Securities and Exchange Commission (``SEC'').  This report, the so-called Form 10-K as defined in sections 13 and 15(d) of the Securities Exchange Act of 1934 (15 U.S.C. § 78o or 78o(d)), provides a broad overview of a company's performance and its risks.  Unlike other sources of information, the statements contained within these reports are certified and attested to by both a company's officers and its independent auditors.  If these statements are negligent or fraudulent, the SEC, Department of Justice, and shareholders are all able to press civil and, in limited cases, criminal charges against officers and auditors.  

The law encourages and sometimes requires disclosure, but the incentives for firms are not entirely one-sided.  Countering the trend towards limitless disclosure is the competition for investor capital.  Through their annual reports, companies generally seek to present themselves as better investment opportunities than their competitors. Their officers thus face a balancing act as between describing a pessimistic future full of potential risks and an optimistic future without any risks.  While certainly not a perfect description of reality, no other source of information, including surveys and press releases, provides as comprehensive of a statement of the environment in which companies operate.

While 10-Ks contain a significant amount of information regarding risks facing the company, we are particularly interested in the regulatory risk and uncertainty highlighted in these filings.  Thus, in this paper, we analyze more than 20 years, 30,000 companies, and 160,000 10-K reports to identify more than 4.5 million references to U.S. Federal regulatory Acts and Agencies.  Using these references, we generate a reproducible, quantitative, and longitudinal measurement of the temperature and diversity of the U.S. regulatory ``ecosystem.''  We document a clear trend towards increasing total energy, temperature, and diversity, with double- to triple-digit growth in all measurements. We believe this framework and its ongoing application represent a principled approach to the quantification of both the global and various local regulatory ecosystems, and we hope that this research can drive better-grounded discussions of legal complexity and policy design in the modern world.

\section{Data}
\label{S:2}	
10-K filings have been the focus of many academic studies in finance and accounting \cite{campbell2014information}, \cite{you2009financial}, \cite{kravet2013textual}, \cite{li2010information}, law \cite{nelson2016carrot}, \cite{doran2009climate} and other adjacent fields \cite{bao2014simultaneously}, \cite{huang2011multilabel}, \cite{tsai2017risk}.  Many of these studies have focused upon questions such as whether the reporting requirements are achieving their intended purpose and the extent to which markets react to the information disclosed in such findings.  While these are certainly worthy questions in their own right, we believe these filings reveal important patterns that, \textit{at scale} and \textit{over time}, provide meaningful insight into a range of other scientific questions. 

Companies that meet the requirements for 10-K reporting expend significant and increasing resources to prepare these documents, typically with the assistance of accounting firms and lawyers.  Indeed, the Annual Audit Fee Survey \cite{ferf2016audit} conducted by the Financial Executives Research Foundation reveals a mean and median 2015 expense of \$1.8M and \$522,205, respectively, across over 6,000 filers. As required by law, the figures and statements contained within these reports are certified and attested to by both a company's officers and its independent auditors. Therefore, unlike other sources of information, in general, these 10-K annual reports are more likely to convey comprehensive and realistic description of the environment in which companies operate.

Form 10-K filings generally contain at least four parts and fifteen schedules, which collectively offer a wealth of useful information about registered companies.  These parts include a characterization of a company's financial health, legal risks, and other systematic and idiosyncratic factors, such as the nature of the regulatory environment in which it operates. Some of these factors, such as tax credits, may be positive, but the majority of listed regulatory factors are present as risks.   While there are a number of specific requirements under the law, firms and industries are provided with some latitude regarding how to satisfy reporting requirements.  In addition, as explored in \cite{campbell2014information} and \cite{nelson2016carrot}, there have been some important changes in the reporting rules over time.  That said, SEC form templates and accounting firm standards result in more similarity than difference across firms and across time.

Through this exercise of risk factor disclosure, companies typically describe various sources of regulatory risk, including the laws and administrative agencies that are most relevant to their respective businesses.  Consider, for example, the 2015 10-K filing of Trans Energy, Inc., an oil and gas exploration company. Among other statutory and agency references, their filing references both the Migratory Bird Treaty Act of 1918 and Endangered Species Act of 1973:
\begin{quotation}
\textbf{The Endangered Species Act} (“ESA”) was established to protect endangered and threatened species. Pursuant to that act, if a species is listed as threatened or endangered, restrictions may be imposed on activities adversely affecting that species’ habitat. Similar protections are offered to migratory birds under the \textbf{Migratory Bird Treaty Act}. The Company conducts operations on oil and natural gas leases that have species, such as raptors, that are listed and species, such as sage grouse, that could be listed as threatened or endangered under the ESA.
\end{quotation}

Across the broader set of required company disclosures, 10-K filings are filled with references such as these; some, such as citations to sections 13 and 15(d) of the Securities Exchange Act of 1934, are boilerplate and required by the SEC's forms.  Others, such as the 756 references to the Migratory Bird Treaty Act of 1918, or the 27 references to the Price-Anderson Nuclear Industries Indemnity Act of 1957 over the last 23 years, are not.

In the mid-1990s, the SEC introduced its Electronic Data Gathering, Analysis, and Retrieval (``EDGAR'') system.  Since then, nearly all registered company 10-K reports have been uploaded and made available on EDGAR, resulting in more than 160,000 10-K reports accessible online.  We retrieve these 10-K reports and build a multi-stage pipeline that identifies and normalizes references to Acts and Agencies. References are first identified through standard natural language processing techniques; once a reference fragment is identified, it is then passed through a second stage of normalization.  As one example, many filers reference the ``Gramm Leach Bliley Financial Services Modernization Act of 1999;'' however, they do not do so using its full name, as above. Instead, they frequently refer to it as ``GLB,'' ``Graham Leach Bliley,'' ``Gramm Leach Bliley,'' or the ``Financial Services Modernization'' Act.  In order to handle this variation, we built a mapping for over 600 potential Act references, relying on a combination of the US Code, Wikipedia, and manual review.  This mapping is then combined with fuzzy-string matching techniques to correct for spelling mistakes such as ``Graham Leach Bliley.''  The result is a high-precision and high-recall extraction of 401 unique Federal Acts and 133 Agencies across our 23-year dataset. In total, we identify more than 4.5 million Act and Agency references contained in 10-K reports over the past 23 years.    

\section{Methodology and Results}
\label{S:3}
Temperature, at its theoretical basis, is a measure of energy per unit of area or volume.  The more energy per unit of space, \textit{c.p.}, the higher the temperature.  While the actual behavior of matter may vary based on whether this energy is vibrational, rotational, or translational, temperature \textit{per se} provides an incredibly valuable mean-field characterization of the behavior thermodynamic system.  

In our context, each reference to an Act or Agency is equivalent to some energy expended by an accountant, attorney, or company to describe some dimension of regulatory exposure.  While not every potential exposure a company faces is listed, and those that are listed do not have an ``equal'' energy that can be measured in $J$ or $W$, these counts overall correspond to the total number and range of regulatory concepts addressed therein.  This therefore allows us to leverage these references to provide a mean-field characterization of the overall regulatory environment inhabited by registered companies. 

Like the global mean temperature of the Earth itself, there are certainly localized ``regulatory micro-climates'' whose temperature differs from the global climate.  We appreciate that these micro-climates are of great importance across many professional and academic questions, but for purposes of this paper, we focus on aggregate, longitudinal patterns in this paper, relying on millions of references in hundreds of thousands of filings for tens of thousands of companies across more than two decades.

To operationalize these ideas, we first define a regulatory ``space'' and filing ``profile.''  The regulatory space consists of 401 dimensions - one for each discrete Act currently identified in our data.  We can represent a filling as a vector in this regulatory space.  In the simplest form, a 10-K profile $\vec{p}(a)$ has element $p_i(a)$ equal to the number of references to Act $i$ for a company $a$'s annual filing.  This vector $\vec{p}$ can then be normalized or projected.  For example, we can project $\vec{p}$ from the number of references to a ``bitstring'' vector $\vec{b}(a)$, whose element $b_i(a)$ is equal to 1 if a company $a$'s filing mentions Act $i$ at least once, else 0.  Alternatively, we can aggregate or normalize these filings by viewing them as a time-indexed collection.  Let $F(t)$ be the matrix whose rows correspond to the set of all $\vec{p}$ or $\vec{b}$ vectors filed within a given year $t$.  Then the grand sum of $F(t)$ is the total number of references, the $j^{th}$ column sum of $F(t)$ is the number of references to the $j^{th}$ Act, and the $i^{th}$ row sum is the number of references per firm in year $t$.  It is possible normalize the number of references per Act to a rate of reference per filing by dividing the column sums of $F(t)$ by the number of rows $m(t)$, which we call $\vec{r}_j(t)$ for $j^{th}$ Act in year $t$.

We then use these representations to measure the total energy, temperature, and diversity of the regulatory ecosystem as follows.  First, we measure the total energy of the regulatory ecosystem using $p$-vectors as:
\begin{equation}
E(t) = \sum_{i} \sum_{j} F_{i,j}(t)
\end{equation}

Figure 1 shows that the total energy, as measured by number of references to Acts per year, has increased substantially in the last 23 years.  In 1996, there were just over 40,000 references to Acts in the nearly 5,000 filings that year; by 2006, these numbers had more than quadrupled to nearly 200,000 references in just over 9,000 filings; and, through three quarters of 2016, these numbers have again increased to an annualized rate of over 300,000 references.

\begin{figure}[ht]
\label{F:1}
\centering\includegraphics[width=1.0\linewidth]{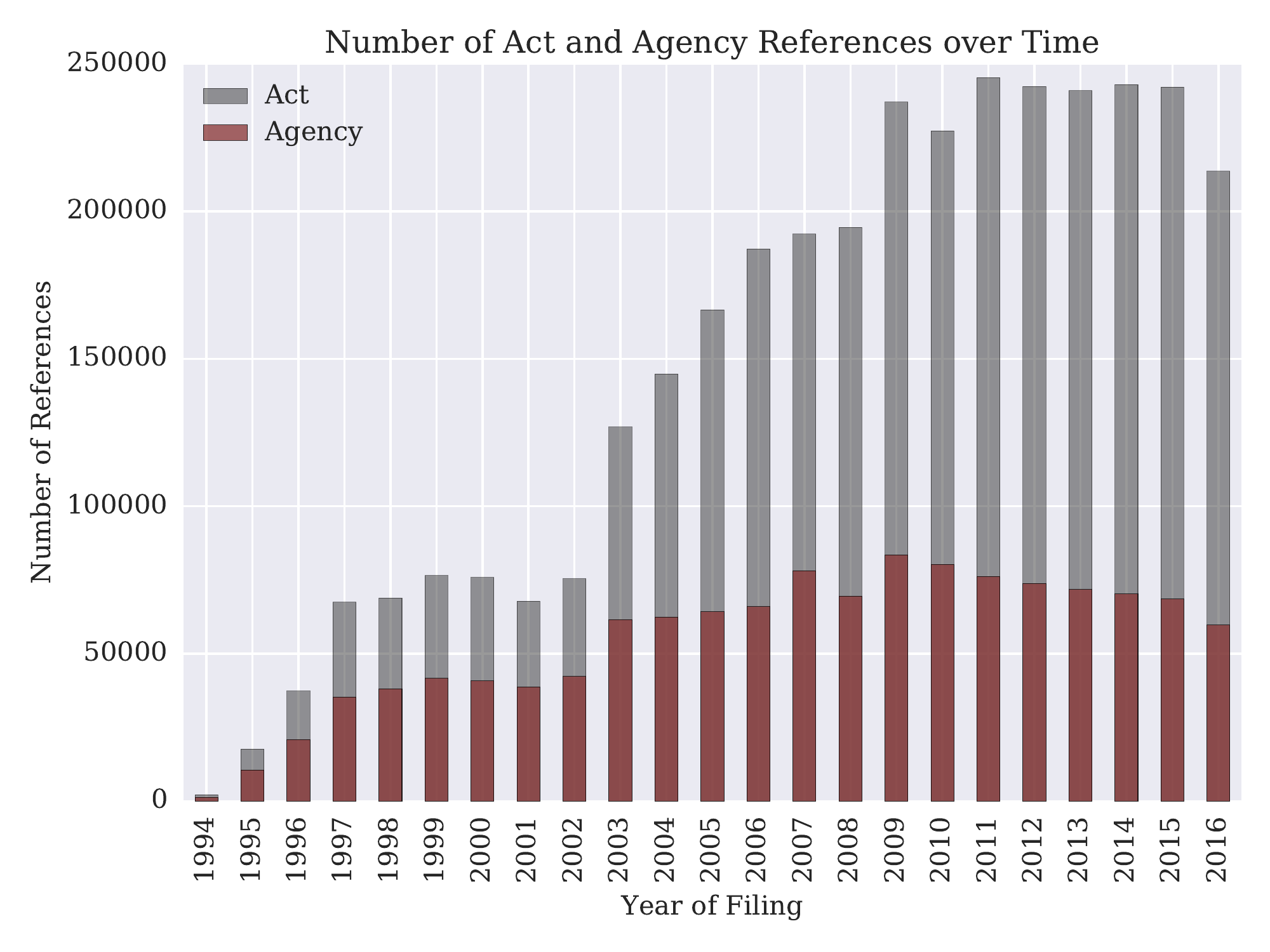}
\caption{Total number of Act references per filing over time.}
\end{figure}

Total energy alone can be misleading with respect to policy interpretations, however, as there are a number of reasons why energy may change without relation to regulatory exposure or ``burden.''  For example, i) the economy may grow or shrink in real or nominal terms, increasing or decreasing the number of companies or companies meeting registration requirement, ii) the SEC rules governing registration or filings may change, increasing or decreasing the number of companies or references, or iii) \textit{c.p.}, the relative incentives to incorporate or take on shareholders may change, increasing or decreasing the number of companies registered.  These factors do not necessarily imply more or less regulation as experienced by individual companies, although they may be viewed as endogenous to some policy questions.

We can control for these factors by normalizing total energy to temperature, taking into account the number of filings per year as an analogy to area or volume.  To do so, we calculate the average rate of references per filing, ``temperature,'' $T(t)$ as follows:
\begin{equation}
T(t) = \frac{E(t)}{m(t)}
\end{equation}

Figure 2 shows that, over the last 23 years, $T(t)$ has been monotonically increasing.  While the rate of reference, like the total energy in Figure 1 above, clearly shows the effect of the Sarbanes-Oxley changes in 2003, this trend remains unbroken both well before and well after.  In 1996, the average number of references per filing was 8.4; by 2006, it had more than doubled to 20.9; and by 2016, the rate had increased again by more than 50\% to 31.7 Act references per filing.  Even if the amount of energy or cost does not scale linearly per filing with the number of references, the monotonic, 237\% increase in $T(t)$ clearly demonstrates an increasing regulatory temperature.

\begin{figure}[ht]
\label{F:2}
\centering\includegraphics[width=1.0\linewidth]{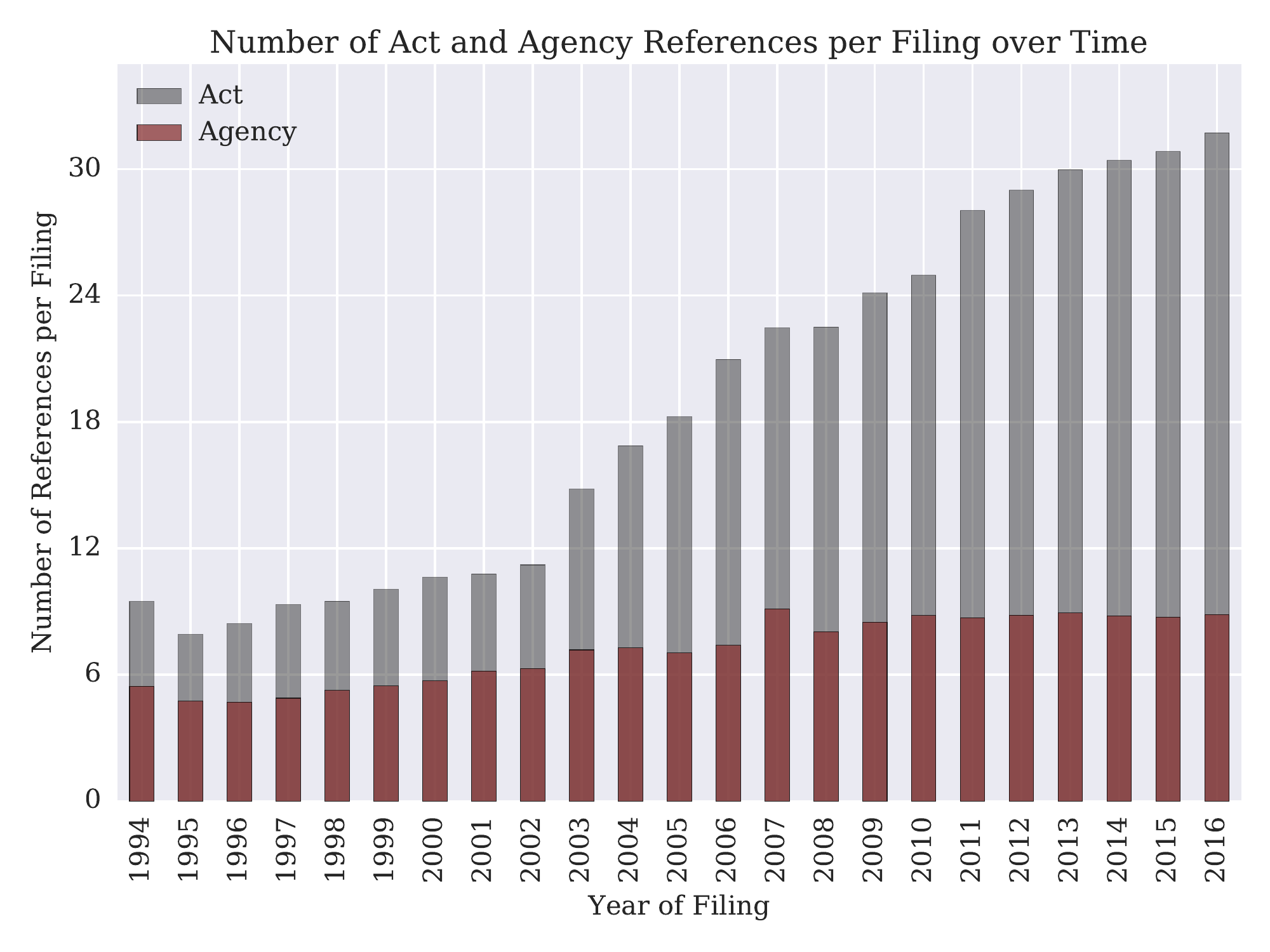}
\caption{Average number of Act references per filing over time.}
\end{figure}

Table 1 below summarizes the data from Figures 1 and 2 above.

\begin{table}[ht]
\begin{center}
\begin{tabular}{|r|r|r|}
\hline
Year & $E(t)$ & $T(t)$\\
\hline
1995       &   17,672 &   7.9 \\\hline
2000       &   75,851 &  10.6 \\\hline
2005       &  166,518 &  18.2 \\\hline
2010       &  227,210 &  25.0 \\\hline
2015       &  242,107 &  30.8 \\\hline
\end{tabular}
\label{T:1}
\caption{Summary of energy and temperature measures over time.}
\end{center}
\end{table}

Finally, we may ask - is temperature or energy changing in concert with diversity, or is the change in temperature concentrated along a single dimension of regulation?  For example, changes in energy or temperature can represent more or less reliance on the same Act, e.g., the Securities Exchange Act of 1934; in this case, the number of unique Acts referenced is not changing, but the regulatory exposure per Act is.  Alternatively, the total number of unique Acts referenced could be increasing or decreasing; for example, in 2003, most registered companies added references to the recently enacted Sarbanes-Oxley Act,  which had not previously been referenced.  Changes such as these represent an increase or decrease in the number of dimensions or diversity of regulatory exposure, but not necessarily the intensity of each exposure.

Using our notation above, we evaluate the diversity question by calculating two measures.  First, we calculate the number of unique Acts per filing through the sum of $b$ vectors above.  Then, we calculate the average number of unique Acts per filing, across all companies in a given year; this is $\frac{1}{m} \sum_{j} \sum_{i} b_i^j$, where $b_i^j$ is the bit corresponding to whether the $i^{th}$ Act was referenced in the $j^{th}$ company filing and $m$ is the number of companies per year.

Figure 3 shows that, over the last 23 years, the diversity of Acts referenced has increased jointly with temperature.  Like Figures 1 and 2 above, the time series exhibits a jump following Sarbanes-Oxley; however, like Figure 2, the time series also exhibits a monotonic increase over two decades, growing from 3.1 unique Acts per filing in 1996 to 5.6 unique Acts per filing in 2006 to 7.9 Acts per filing in 2016.  This increase suggests that the increase in regulatory ecosystem temperature has been, at least in part, related to an increase in the number of regulatory ``micro-climates'' present in the global regulatory ecosystem.

\begin{figure}[ht]
\label{F:3}
\centering\includegraphics[width=1.\linewidth]{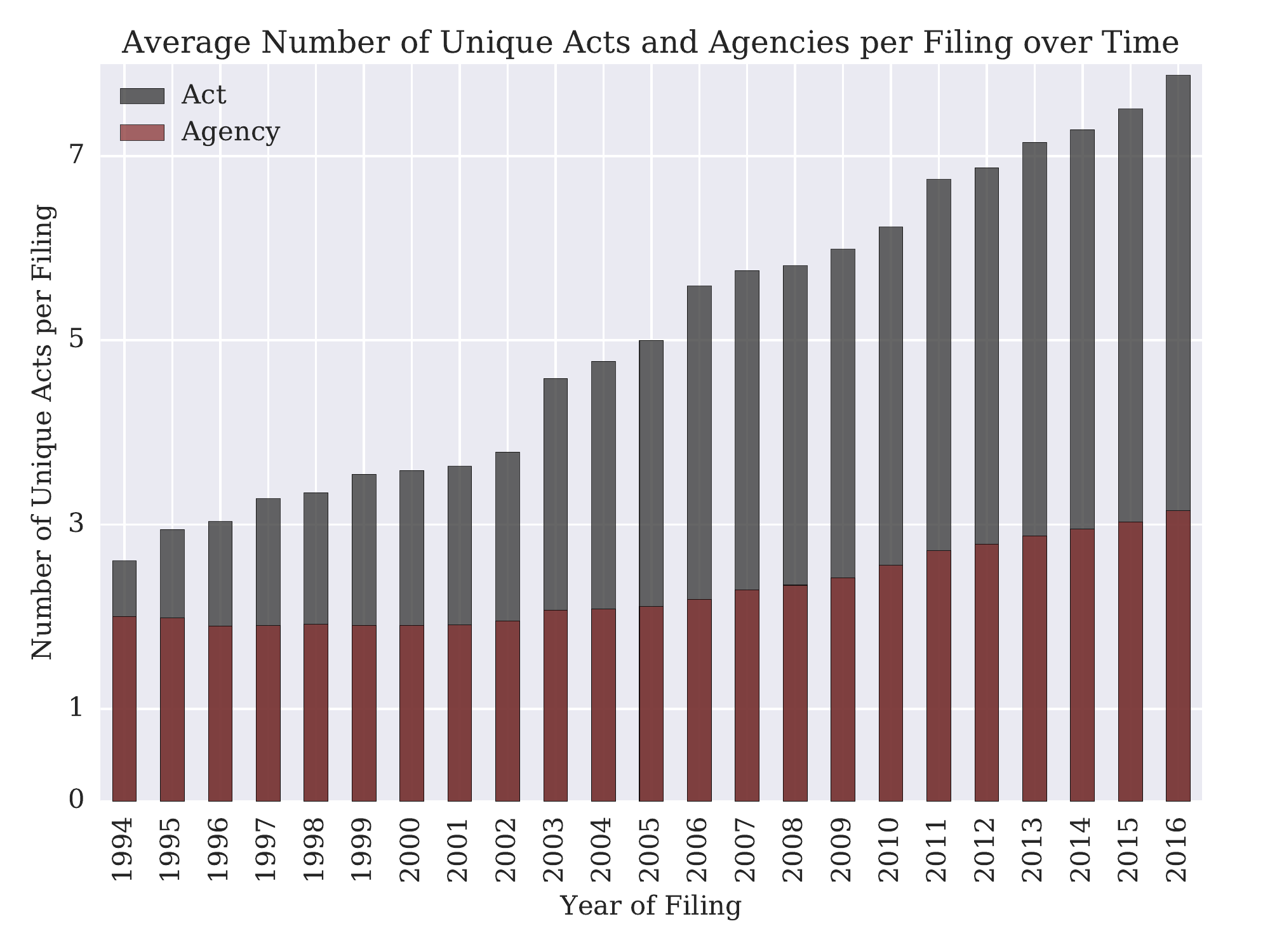}
\caption{Average number of unique Acts and Agencies per filing over time.}
\end{figure}

As an additional measure of diversity, we analyze each company's yearly regulatory ``bitstring.''  As noted earlier, we calculate the 401-bit vector $b$ for each company-year, where each bit corresponds to the presence of the 401 discrete Acts we identify.  Although the regulatory space has 401 dimensions, the bitstring for a given filing is likely to be extremely sparse. For example, consider the 2012 10-K filed by the \textit{Boeing Company}. Their filing features a bitstring with 12 non-zero elements, including Acts such as the Homeland Security Act, the Employee Retirement Income Security Act, the Patient Protection And Affordable Care Act and the American Taxpayer Relief.  Alternatively, the 2014 10-K of \textit{Facebook Inc.} features 10 unique elements, including The Bank Secrecy Act, the U.S. Foreign Corrupt Practices Act, the USA Patriot Act, and the Credit CARD Act.

After applying this formalization to all companies for all years,  we calculate the average pairwise Hamming distance \cite{hamming1950error} between all company bitstrings in a given year. Hamming distance is commonly used to evaluate the diversity of genomic \cite{tully2016differences}, \cite{he2004genetic}, \cite{mattiussi2004measures} and other related data \cite{martin2015euclidean}, \cite{alfred2014algorithms}, \cite{morrison2001measurement}. It can be interpreted as proportional to the average number of regulatory dimensions not in common between companies.  More explicitly, the Hamming distance between two companies $a$ and $b$ in year $t$ is:
\begin{equation}
d_{a,b}(t) = \sum_{i} \vec{p}(a) \oplus \vec{p}(b)
\end{equation}
where $\oplus$ is the element-wise XOR operator.  We can then write the average Hamming distance $\bar{d}(t)$ as the average over all combinations of $a$ and $b$ at $t$.  Figure 4 visualizes the structure of the distance matrix $D$ for all $a,b$ as of 1994.  The large block in the lower right corresponds primarily to special purpose vehicles like trusts or limited partnerships, and the overall structure corresponds to sectors and industries.

\begin{figure}[ht]
\label{F:4}
\centering\includegraphics[width=1.0\linewidth]{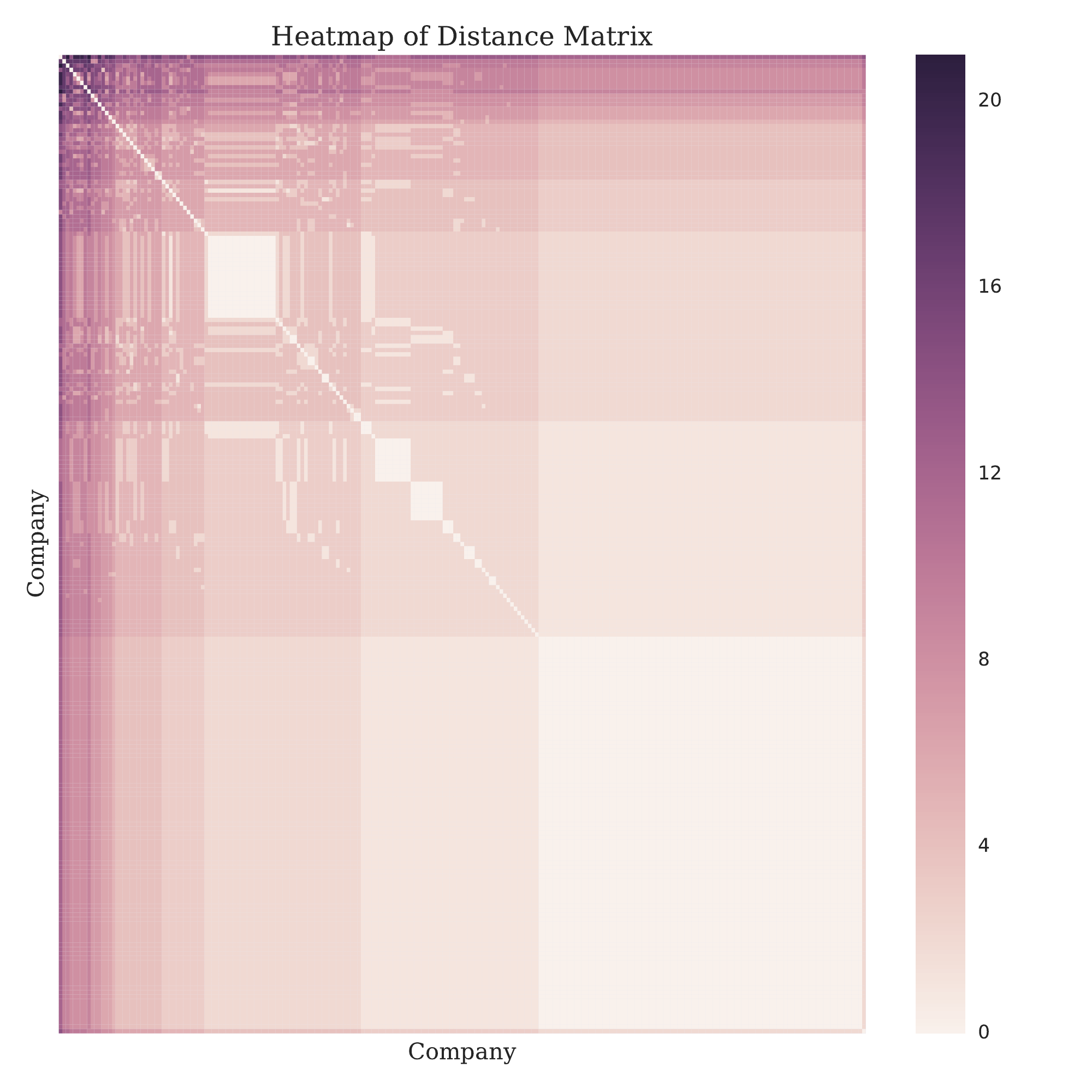}
\caption{Hamming distance heatmap for 1994 Act bitstrings.}
\end{figure}

Figure 5, the average Hamming distance $\bar{d}(t)$ over time for Acts and Agencies, portrays mean-field distance between firms at scale, confirming an increasing diversity across the global regulatory ecosystem.  Over time, companies are subject, on average, to increasingly different requirements.  While not monotonically increasing like the rate of reference and number of unique references above, the average distance increases 18 of 23 years in the sample.  In 1996, two firms were separated on average by fewer than four regulatory Act ``bits'' or ``genes''; by 2016, this number has increased to nearly 10.

\begin{figure}[ht]
\label{F:5}
\centering\includegraphics[width=1.0\linewidth]{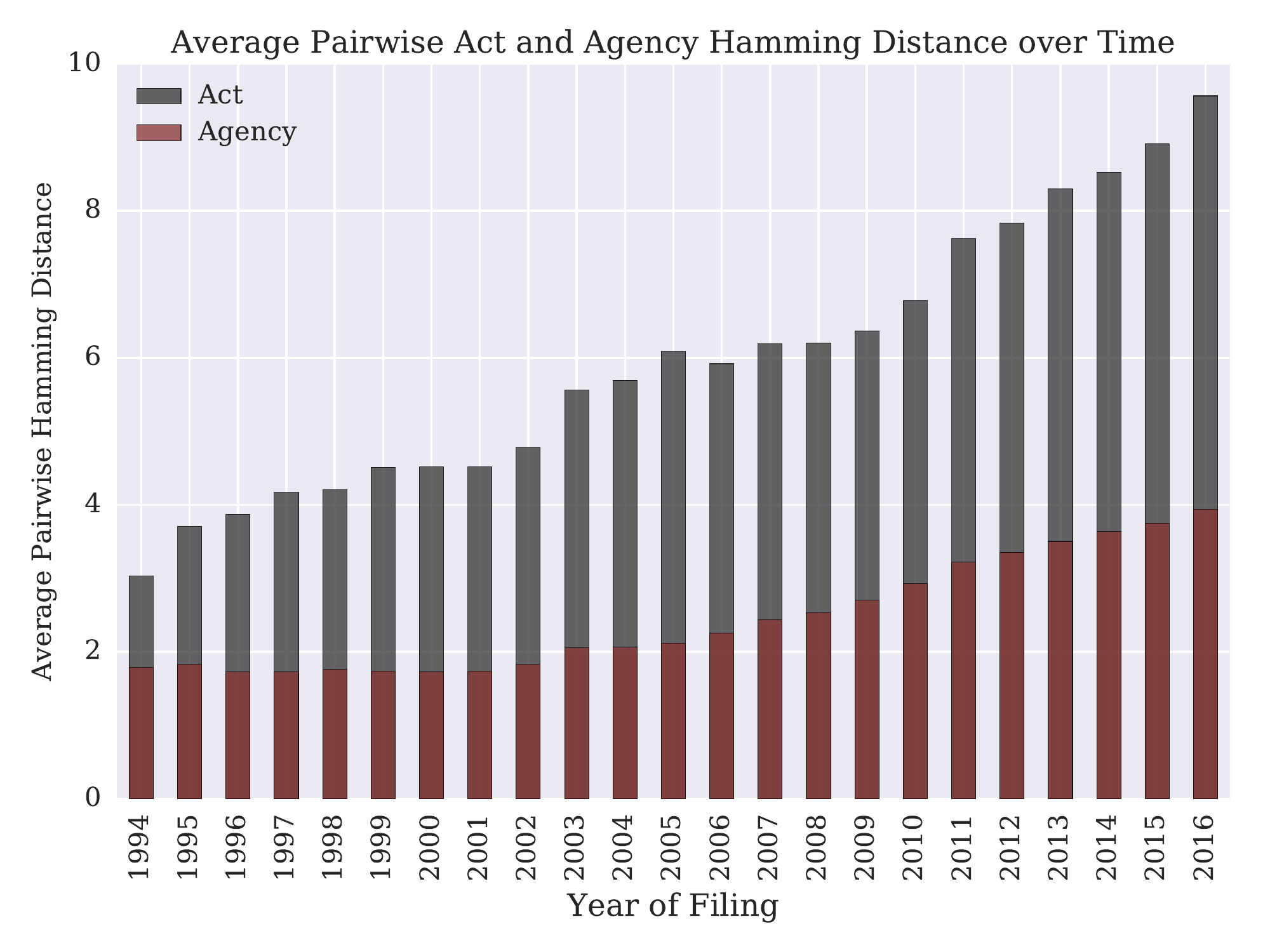}
\caption{Average pairwise Hamming distance between company filing regulatory bitstrings over time.}
\end{figure}

Table 2 summarizes the data from Figures 3 and 4 above.

\begin{table}[ht]
\begin{center}
\begin{tabular}{|r|r|r|}
\hline
Year & Average Unique Acts & $\bar{d}(t)$\\
\hline
1995       &  2.9 &       3.7 \\\hline
2000       &  3.6 &       4.5 \\\hline
2005       &  5.0 &       6.1 \\\hline
2010       &  6.2 &       6.8 \\\hline
2015       &  7.5 &       8.9 \\
\hline
\end{tabular}
\label{T:2}
\caption{Summary of diversity measures over time.}
\end{center}
\end{table}

\section{Conclusion and Future Work}
\label{S:4}
In this paper, we have presented the first large-scale, longitudinal characterization of the energy, temperature, and diversity of the regulatory ecosystem.  We have identified increasing regulatory exposure along an increasing number of dimensions, providing evidence in support of the claim that regulatory burden is increasing.  Using a bitstring representation of firm regulatory exposure, we have confirmed that the aggregate Federal regulatory ecosystem is becoming more diverse over time, providing evidence in support of the claim that regulatory complexity is increasing.  These conclusions are based on more than 20 years, 30,000 companies, 160,000 10-K reports, and 4.5 million references contained in uniquely comprehensive and accurate 10-K reports.
 
In future work, we intend to expand upon these questions and connect to extant research agenda, including the categorization of regulatory ``species'' and ``climates'' through their time series signatures and linguistic markers, a mapping of the time-dependent, tri-partite network of firms, Acts, and Agencies, and the integration of this analysis with our existing work on the complexity of other US statutory, regulatory, and judicial systems \cite{bommarito2010mathematical}, \cite{katz2014measuring}.

Our work contributes to both the broader literature on legal complexity \cite{ruhl2015measuring}, \cite{boulet2010network}, \cite{bourcier2007toward} and efforts to document the physical properties of legal systems as complex adaptive systems \cite{lee2015statistical}, \cite{clark2012genealogy}, \cite{katz2010hustle}, \cite{smith2007web}, \cite{jones2007dynamical}, \cite{post2000long}.  In addition, this paper is among a growing set of recent works applying tools of machine learning and natural language processing to better understand the behavior of various legal systems \cite{aletras2016predicting}, \cite{katz2016predicting}, \cite{mcshane2012predicting}.

In sum, we believe that this framework for representation and measurement will contribute to ongoing academic and policy discussions around legal complexity and policy design. The continued development of both global and local regulatory indices can provide for a principled, empirical basis of evaluation, standing in stark contrast to some of the vague generalizations that frequently guide current discourse.
\\
\newpage

\bibliographystyle{model1-num-names}
\bibliography{measuring_regulatory_ecosystem.bib}

\begin{thebibliography}{31}
\expandafter\ifx\csname natexlab\endcsname\relax\def\natexlab#1{#1}\fi
\providecommand{\bibinfo}[2]{#2}
\ifx\xfnm\relax \def\xfnm[#1]{\unskip,\space#1}\fi
\bibitem[{Campbell et~al.(2014)Campbell, Chen, Dhaliwal, Lu, and
  Steele}]{campbell2014information}
\bibinfo{author}{J.~L. Campbell}, \bibinfo{author}{H.~Chen},
  \bibinfo{author}{D.~S. Dhaliwal}, \bibinfo{author}{H.-m. Lu},
  \bibinfo{author}{L.~B. Steele},
\newblock \bibinfo{title}{The information content of mandatory risk factor
  disclosures in corporate filings},
\newblock \bibinfo{journal}{Review of Accounting Studies} \bibinfo{volume}{19}
  (\bibinfo{year}{2014}) \bibinfo{pages}{396--455}.
\bibitem[{You and Zhang(2009)}]{you2009financial}
\bibinfo{author}{H.~You}, \bibinfo{author}{X.-j. Zhang},
\newblock \bibinfo{title}{Financial reporting complexity and investor
  underreaction to 10-k information},
\newblock \bibinfo{journal}{Review of Accounting Studies} \bibinfo{volume}{14}
  (\bibinfo{year}{2009}) \bibinfo{pages}{559--586}.
\bibitem[{Kravet and Muslu(2013)}]{kravet2013textual}
\bibinfo{author}{T.~Kravet}, \bibinfo{author}{V.~Muslu},
\newblock \bibinfo{title}{Textual risk disclosures and investors’ risk
  perceptions},
\newblock \bibinfo{journal}{Review of Accounting Studies} \bibinfo{volume}{18}
  (\bibinfo{year}{2013}) \bibinfo{pages}{1088--1122}.
\bibitem[{Li(2010)}]{li2010information}
\bibinfo{author}{F.~Li},
\newblock \bibinfo{title}{The information content of forward-looking statements
  in corporate filings—a na{\"\i}ve bayesian machine learning approach},
\newblock \bibinfo{journal}{Journal of Accounting Research}
  \bibinfo{volume}{48} (\bibinfo{year}{2010}) \bibinfo{pages}{1049--1102}.
\bibitem[{Nelson and Pritchard(2016)}]{nelson2016carrot}
\bibinfo{author}{K.~K. Nelson}, \bibinfo{author}{A.~C. Pritchard},
\newblock \bibinfo{title}{Carrot or stick? the shift from voluntary to
  mandatory disclosure of risk factors},
\newblock \bibinfo{journal}{Journal of Empirical Legal Studies}
  \bibinfo{volume}{13} (\bibinfo{year}{2016}) \bibinfo{pages}{266--297}.
\bibitem[{Doran and Quinn(2009)}]{doran2009climate}
\bibinfo{author}{K.~L. Doran}, \bibinfo{author}{E.~L. Quinn},
\newblock \bibinfo{title}{Climate change risk disclosure: a sector by sector
  analysis of sec 10-k filings from 1995-2008},
\newblock \bibinfo{journal}{North Carolina Journal of International Law and
  Commercial Regulation} \bibinfo{volume}{34} (\bibinfo{year}{2009})
  \bibinfo{pages}{101--147}.
\bibitem[{Bao and Datta(2014)}]{bao2014simultaneously}
\bibinfo{author}{Y.~Bao}, \bibinfo{author}{A.~Datta},
\newblock \bibinfo{title}{Simultaneously discovering and quantifying risk types
  from textual risk disclosures},
\newblock \bibinfo{journal}{Management Science} \bibinfo{volume}{60}
  (\bibinfo{year}{2014}) \bibinfo{pages}{1371--1391}.
\bibitem[{Huang and Li(2011)}]{huang2011multilabel}
\bibinfo{author}{K.-W. Huang}, \bibinfo{author}{Z.~Li},
\newblock \bibinfo{title}{A multilabel text classification algorithm for
  labeling risk factors in sec form 10-k},
\newblock \bibinfo{journal}{ACM Transactions on Management Information Systems
  (TMIS)} \bibinfo{volume}{2} (\bibinfo{year}{2011}) \bibinfo{pages}{1--19}.
\bibitem[{Tsai and Wang(2017)}]{tsai2017risk}
\bibinfo{author}{M.-F. Tsai}, \bibinfo{author}{C.-J. Wang},
\newblock \bibinfo{title}{On the risk prediction and analysis of soft
  information in finance reports},
\newblock \bibinfo{journal}{European Journal of Operational Research}
  \bibinfo{volume}{257} (\bibinfo{year}{2017}) \bibinfo{pages}{243--250}.
\bibitem[{Foundation(2016)}]{ferf2016audit}
\bibinfo{author}{F.~E.~R. Foundation}, \bibinfo{title}{2016 audit fee report},
  \bibinfo{year}{2016}.
\bibitem[{Hamming(1950)}]{hamming1950error}
\bibinfo{author}{R.~W. Hamming},
\newblock \bibinfo{title}{Error detecting and error correcting codes},
\newblock \bibinfo{journal}{Bell System technical journal} \bibinfo{volume}{29}
  (\bibinfo{year}{1950}) \bibinfo{pages}{147--160}.
\bibitem[{Tully et~al.(2016)Tully, Ogilvie, Batorsky, Bean, Power,
  Ghebremichael, Bedard, Gladden, Seese, Amero et~al.}]{tully2016differences}
\bibinfo{author}{D.~C. Tully}, \bibinfo{author}{C.~B. Ogilvie},
  \bibinfo{author}{R.~E. Batorsky}, \bibinfo{author}{D.~J. Bean},
  \bibinfo{author}{K.~A. Power}, \bibinfo{author}{M.~Ghebremichael},
  \bibinfo{author}{H.~E. Bedard}, \bibinfo{author}{A.~D. Gladden},
  \bibinfo{author}{A.~M. Seese}, \bibinfo{author}{M.~A. Amero}, et~al.,
\newblock \bibinfo{title}{Differences in the selection bottleneck between modes
  of sexual transmission influence the genetic composition of the hiv-1 founder
  virus},
\newblock \bibinfo{journal}{PLoS Pathog} \bibinfo{volume}{12}
  (\bibinfo{year}{2016}) \bibinfo{pages}{e1005619}.
\bibitem[{He et~al.(2004)He, Petoukhov, and Ricci}]{he2004genetic}
\bibinfo{author}{M.~X. He}, \bibinfo{author}{S.~V. Petoukhov},
  \bibinfo{author}{P.~E. Ricci},
\newblock \bibinfo{title}{Genetic code, hamming distance and stochastic
  matrices},
\newblock \bibinfo{journal}{Bulletin of mathematical biology}
  \bibinfo{volume}{66} (\bibinfo{year}{2004}) \bibinfo{pages}{1405--1421}.
\bibitem[{Mattiussi et~al.(2004)Mattiussi, Waibel, and
  Floreano}]{mattiussi2004measures}
\bibinfo{author}{C.~Mattiussi}, \bibinfo{author}{M.~Waibel},
  \bibinfo{author}{D.~Floreano},
\newblock \bibinfo{title}{Measures of diversity for populations and distances
  between individuals with highly reorganizable genomes},
\newblock \bibinfo{journal}{Evolutionary Computation} \bibinfo{volume}{12}
  (\bibinfo{year}{2004}) \bibinfo{pages}{495--515}.
\bibitem[{Martin and Cao(2015)}]{martin2015euclidean}
\bibinfo{author}{E.~Martin}, \bibinfo{author}{E.~Cao},
\newblock \bibinfo{title}{Euclidean chemical spaces from molecular
  fingerprints: Hamming distance and hempel’s ravens},
\newblock \bibinfo{journal}{Journal of computer-aided molecular design}
  \bibinfo{volume}{29} (\bibinfo{year}{2015}) \bibinfo{pages}{387--395}.
\bibitem[{Alfred(2014)}]{alfred2014algorithms}
\bibinfo{author}{V.~Alfred}, \bibinfo{title}{Algorithms for finding patterns in
  strings}, \bibinfo{publisher}{Elsevier}, \bibinfo{year}{2014}.
\bibitem[{Morrison and De~Jong(2001)}]{morrison2001measurement}
\bibinfo{author}{R.~W. Morrison}, \bibinfo{author}{K.~A. De~Jong},
\newblock \bibinfo{title}{Measurement of population diversity},
\newblock in: \bibinfo{booktitle}{International Conference on Artificial
  Evolution (Evolution Artificielle)}, \bibinfo{organization}{Springer}, pp.
  \bibinfo{pages}{31--41}.
\bibitem[{Bommarito and Katz(2010)}]{bommarito2010mathematical}
\bibinfo{author}{M.~J. Bommarito}, \bibinfo{author}{D.~M. Katz},
\newblock \bibinfo{title}{A mathematical approach to the study of the united
  states code},
\newblock \bibinfo{journal}{Physica A: Statistical Mechanics and its
  Applications} \bibinfo{volume}{389} (\bibinfo{year}{2010})
  \bibinfo{pages}{4195--4200}.
\bibitem[{Katz and Bommarito~II(2014)}]{katz2014measuring}
\bibinfo{author}{D.~M. Katz}, \bibinfo{author}{M.~J. Bommarito~II},
\newblock \bibinfo{title}{Measuring the complexity of the law: the united
  states code},
\newblock \bibinfo{journal}{Artificial Intelligence and Law}
  \bibinfo{volume}{22} (\bibinfo{year}{2014}) \bibinfo{pages}{337--374}.
\bibitem[{Ruhl and Katz(2015)}]{ruhl2015measuring}
\bibinfo{author}{J.~Ruhl}, \bibinfo{author}{D.~M. Katz},
\newblock \bibinfo{title}{Measuring, monitoring, and managing legal
  complexity},
\newblock \bibinfo{journal}{Iowa Law Review} \bibinfo{volume}{101}
  (\bibinfo{year}{2015}) \bibinfo{pages}{191---244}.
\bibitem[{Boulet et~al.(2010)Boulet, Mazzega, and Bourcier}]{boulet2010network}
\bibinfo{author}{R.~Boulet}, \bibinfo{author}{P.~Mazzega},
  \bibinfo{author}{D.~Bourcier},
\newblock \bibinfo{title}{Network analysis of the french environmental code},
\newblock in: \bibinfo{booktitle}{Ai approaches to the complexity of legal
  systems. complex systems, the semantic web, ontologies, argumentation, and
  dialogue}, \bibinfo{publisher}{Springer}, \bibinfo{year}{2010}, pp.
  \bibinfo{pages}{39--53}.
\bibitem[{Bourcier and Mazzega(2007)}]{bourcier2007toward}
\bibinfo{author}{D.~Bourcier}, \bibinfo{author}{P.~Mazzega},
\newblock \bibinfo{title}{Toward measures of complexity in legal systems},
\newblock in: \bibinfo{booktitle}{Proceedings of the 11th international
  conference on Artificial intelligence and law}, \bibinfo{organization}{ACM},
  pp. \bibinfo{pages}{211--215}.
\bibitem[{Lee et~al.(2015)Lee, Broedersz, and Bialek}]{lee2015statistical}
\bibinfo{author}{E.~D. Lee}, \bibinfo{author}{C.~P. Broedersz},
  \bibinfo{author}{W.~Bialek},
\newblock \bibinfo{title}{Statistical mechanics of the us supreme court},
\newblock \bibinfo{journal}{Journal of Statistical Physics}
  (\bibinfo{year}{2015}) \bibinfo{pages}{1--27}.
\bibitem[{Clark and Lauderdale(2012)}]{clark2012genealogy}
\bibinfo{author}{T.~S. Clark}, \bibinfo{author}{B.~E. Lauderdale},
\newblock \bibinfo{title}{The genealogy of law},
\newblock \bibinfo{journal}{Political Analysis} \bibinfo{volume}{20}
  (\bibinfo{year}{2012}) \bibinfo{pages}{329--350}.
\bibitem[{Katz and Stafford(2010)}]{katz2010hustle}
\bibinfo{author}{D.~M. Katz}, \bibinfo{author}{D.~K. Stafford},
\newblock \bibinfo{title}{Hustle and flow: A social network analysis of the
  american federal judiciary},
\newblock \bibinfo{journal}{Ohio State Law Journal} \bibinfo{volume}{71}
  (\bibinfo{year}{2010}) \bibinfo{pages}{457--507}.
\bibitem[{Smith(2007)}]{smith2007web}
\bibinfo{author}{T.~A. Smith},
\newblock \bibinfo{title}{The web of the law},
\newblock \bibinfo{journal}{San Diego L. Rev.} \bibinfo{volume}{44}
  (\bibinfo{year}{2007}) \bibinfo{pages}{309--354}.
\bibitem[{Jones(2007)}]{jones2007dynamical}
\bibinfo{author}{G.~T. Jones},
\newblock \bibinfo{title}{Dynamical jurisprudence: law as a complex system},
\newblock \bibinfo{journal}{Ga. St. UL Rev.} \bibinfo{volume}{24}
  (\bibinfo{year}{2007}) \bibinfo{pages}{873--883}.
\bibitem[{Post and Eisen(2000)}]{post2000long}
\bibinfo{author}{D.~G. Post}, \bibinfo{author}{M.~B. Eisen},
\newblock \bibinfo{title}{How long is the coastline of law? thoughts on the
  fractal nature of legal systems},
\newblock \bibinfo{journal}{Journal of Legal Studies} \bibinfo{volume}{29}
  (\bibinfo{year}{2000}) \bibinfo{pages}{545--584}.
\bibitem[{Aletras et~al.(2016)Aletras, Tsarapatsanis, Preo{\c{t}}iuc-Pietro,
  and Lampos}]{aletras2016predicting}
\bibinfo{author}{N.~Aletras}, \bibinfo{author}{D.~Tsarapatsanis},
  \bibinfo{author}{D.~Preo{\c{t}}iuc-Pietro}, \bibinfo{author}{V.~Lampos},
\newblock \bibinfo{title}{Predicting judicial decisions of the european court
  of human rights: A natural language processing perspective},
\newblock \bibinfo{journal}{PeerJ Computer Science} \bibinfo{volume}{2}
  (\bibinfo{year}{2016}) \bibinfo{pages}{e93}.
\bibitem[{Katz et~al.(2016)Katz, Bommarito, and Blackman}]{katz2016predicting}
\bibinfo{author}{D.~M. Katz}, \bibinfo{author}{M.~J. Bommarito},
  \bibinfo{author}{J.~Blackman},
\newblock \bibinfo{title}{A general approach for predicting the behavior of the
  supreme court of the united states},
\newblock \bibinfo{journal}{arXiv preprint arXiv:1612.03473}
  (\bibinfo{year}{2016}).
\bibitem[{McShane et~al.(2012)McShane, Watson, Baker, and
  Griffith}]{mcshane2012predicting}
\bibinfo{author}{B.~B. McShane}, \bibinfo{author}{O.~P. Watson},
  \bibinfo{author}{T.~Baker}, \bibinfo{author}{S.~J. Griffith},
\newblock \bibinfo{title}{Predicting securities fraud settlements and amounts:
  a hierarchical bayesian model of federal securities class action lawsuits},
\newblock \bibinfo{journal}{Journal of Empirical Legal Studies}
  \bibinfo{volume}{9} (\bibinfo{year}{2012}) \bibinfo{pages}{482--510}.

\end{thebibliography}
\end{document}